\documentclass[lettersize,journal]{IEEEtran}
\usepackage{amsmath,amsfonts}
\usepackage{algorithmic}
\usepackage{algorithm}
\usepackage{array}
\usepackage[caption=false,font=normalsize,labelfont=sf,textfont=sf]{subfig}
\usepackage{textcomp}
\usepackage{stfloats}
\usepackage{url}
\usepackage{verbatim}
\usepackage{graphicx}
\usepackage{siunitx}
\usepackage{cite}
\usepackage{amssymb}
\usepackage[
CJKbookmarks=false,
colorlinks,
breaklinks=true,
linkcolor=blue,
anchorcolor=blue,
citecolor=blue,
urlcolor=blue,
]{hyperref}

\begin{document}

\title{High-Resolution Quantum Sensing with Rydberg Atomic Receiver: Principles, Experiments and Future Prospects}


\author{Minze~Chen,
      Tianqi~Mao,~\IEEEmembership{Member,~IEEE,}
      Zhiao~Zhu,
      Haonan~Feng,\\
      Ge~Gao,
      Zhonghuai~Wu,
      Wei~Xiao,
      Zhongxiang~Li,
      Dezhi~Zheng,


\thanks{Minze Chen, Tianqi Mao, Zhiao Zhu, Haonan Feng, Ge Gao, Zhonghuai Wu, Wei Xiao, Zhongxiang Li and Dezhi Zheng are with Beijing Institute of Technology. The corresponding authors are Dezhi Zheng and Tianqi Mao.}%
}

\maketitle

\begin{abstract}

Quantum sensing using Rydberg atoms offers unprecedented opportunities for next-generation radar systems, transcending classical limitations in miniaturization and spectral agility. Implementing this paradigm for radar sensing, this work proposes a quantum-enhanced radar reception architecture enabled by the emerging Rydberg atomic receiver, replacing conventional antenna-to-mixer chains with a centimeter-scale vapor cell as a quantum RF field sensor and mixer. The proposed approach is based on electromagnetically induced transparency with the Autler-Townes splitting enabling direct RF-to-optical downconversion within the atomic medium via an external co-frequency reference. To circumvent the intrinsic bottleneck on instantaneous bandwidth of atomic receiver, we invoke a non-uniform stepped-frequency synthesis strategy through hybrid manipulation of Rydberg energy states, which combines coarse laser frequency tuning with fine AC-Stark shift compensation. Additionally, we establish a nonlinear response model of the Rydberg atomic homodyne receiver and propose a customized nonlinear compensation method that extends the linear dynamic range by over 7 dB. We develop a compressive sensing algorithm (CS-Rydberg) to suppress noise and mitigate the undersampling problem. Experimentally, we demonstrate a compact prototype achieving centimeter-level ranging precision (RMSE = 1.06 cm) within 1.6–1.9 m. By synthesizing GHz-bandwidth (2.6–3.6 GHz), resolvable target separations down to 15 cm are observed under controlled sparse scenarios. These results not only validate the feasibility of quantum sensing based on Rydberg atomic receivers but also underscore the architecture’s inherent scalability: by harnessing the atoms’ ultra-broad spectral response, the synthesized bandwidth can be extended well beyond the current range, enabling sub-centimeter resolution in future radar systems while preserving quantum-traceable calibration and a highly simplified front end.

\end{abstract}


\section{Introduction}

Quantum sensing has become a transformative paradigm in precision measurement, exploiting quantum coherence to surpass classical detection limits \cite{degen2017quantum,pirandola2018advances}. When deployed for electromagnetic field detection, this approach enables microwave electrometry with exceptional sensitivity and spectral coverage—capabilities critically essential for next-generation radar systems \cite{maccone2020quantum, assouly2023quantum}. Conventional radar architectures, however, face fundamental trade-offs among miniaturization, sensitivity, and bandwidth. Frequency agility necessitates multiple discrete receiver front-ends and processing modules for separate bands, significantly increasing system complexity. High-sensitivity reception requires large-aperture antennas while remaining constrained by the thermal noise floor \cite{wiesbeck2014radar}. These limitations become acutely constraining for ranging applications, which are the cornerstone of modern remote sensing and Earth observation. The key reason is that the achievable ranging resolution is fundamentally governed by the system bandwidth. Furthermore, deployment on space- and load-constrained platforms like satellites imposes stringent demands for integration density and sensitivity \cite{xiao2022near}. It is therefore imperative to explore novel radar receiving architectures capable of overcoming these fundamental constraints \cite{hu2025distributed}.

Rydberg atomic microwave electrometry, a new breakthrough in quantum sensing, provides a new paradigm for electromagnetic field sensing \cite{sedlacek2012microwave}. Rydberg atomic sensors utilize alkali metal atoms (e.g., cesium or rubidium) excited to highly sensitive Rydberg states via coherent optical excitation. Such sensors have been experimentally verified to have sensitivities on the order of $\text{nV}{\cdot}\text{cm}^{-1}{\cdot}\text{Hz}^{-1/2}$ \cite{jing2020atomic, borowka2024continuous,tu2024approaching}. Theoretically, there is at least another order of magnitude of expansion capability to exceed the thermal noise limit of conventional receivers. The response frequency range can be tuned from DC to terahertz without any device replacement. By means of the intrinsic parameters of the atoms, Rydberg atom sensing can be self-calibrated. At the same time, it can be compactly integrated in a vapor cell of less than a centimeter. Moreover, unlike high-frequency classical antennas which inherently exhibit pronounced directionality, Rydberg-based sensors maintain an effectively omnidirectional response across the entire electromagnetic spectrum \cite{yuan2024isotropic}. The core detection mechanism—based on electromagnetically induced transparency (EIT) and Autler-Townes (AT) splitting—provides simultaneous Radio Frequency (RF) -to-optical conversion and coherent mixing without the typical electronic noise and distortion of traditional receivers. It enables integrated measurement capabilities for intensity quantification \cite{sedlacek2012microwave}, precise frequency discrimination \cite{sedlacek2013atom}, polarization analysis \cite{wang2023precise, bao2016polarization}, and phase-sensitive detection \cite{jing2020atomic, simons2019rydberg, berweger2023closed}, all within a unified, coherent physical framework.  As a result, the Rydberg atomic sensor has attracted significant attention as a key technology to revolutionize conventional microwave electrometry in terms of sensitivity, size and spectral agility.

Research in remote sensing with Rydberg atomic receivers is progressing. Experimental validations to date have demonstrated atomic sensing capabilities for moving target detection \cite{zhang2023quantum}, angle-of-arrival determination \cite{robinson2021determining}, multi-base station positioning \cite{yan2023three}, digital beamforming \cite{mao2023digital} and communication under dynamic radar interference \cite{gao2025rydberg}, but significant hurdles persist in their practical deployment for radar ranging tasks. Although the reception frequency of Rydberg atoms is continuously tunable across the full band, the intrinsic response bandwidth within a single EIT window is typically below 10 MHz, fundamentally limited by the atomic lifetime constraint \cite{bohaichuk2022origins, meyer2023simultaneous}. No studies have yet reported methods that are expected to break through this limitation. This limits the theoretical radar resolution to the order of about 15 meters, far from meeting the basic requirements for applications \cite{arik2010collaborative}. While there have been studies exploring continuous frequency detection capabilities under external electromagnetic fields \cite{prajapati2022tv, sapiro2020time}, current methodologies suffer from marked sensitivity deterioration at frequencies offset from atomic resonance. A recent study proposes a theoretical framework of Rydberg atomic radar ranging using linear frequency modulation (LFM) signals \cite{cui2025realizing}. However, the chirp bandwidth used is still limited to a narrow EIT response window, which is far from meeting the resolution and sensitivity requirements of practical radar applications. In addition, the engineering practicability of the study is yet to be further demonstrated. To our knowledge, comprehensive experimental verification of Rydberg-based radar ranging has not yet been reported. The architectural design, engineering implementation, response modeling, and data processing algorithms of the Rydberg Atomic Radar remain research gaps.

Motivated by these considerations, this paper introduces a novel radar architecture that integrates Rydberg atomic homodyne detection with a stepped-frequency waveform strategy, aiming to leverage quantum sensitivity advantages while simultaneously overcoming the inherent resolution limitations. The principal contributions of this study are summarized as follows:

\begin{itemize}

\item Development and implementation of a novel radar receiver paradigm based on Rydberg atomic quantum sensing. By leveraging laser frequency tuning and AC-Stark-effect-based fine adjustment, the atomic reception mechanism enables the detection of stepped frequency continuous waveform (SFCW) signals, facilitating synthesized wideband sensing from discrete frequency points. This framework is inherently scalable to higher bandwidths and increased frequency-domain sampling density.
\item Establishment of an analytical nonlinear response model tailored specifically for Rydberg atomic homodyne receivers, and proposal of a dedicated nonlinear compensation technique improving linear dynamic range by 7 dB.
\item Introduction of a dedicated compressive-sensing-based algorithm (CS-Rydberg), specifically addressing undersampling and sensitivity challenges unique to Rydberg atomic radar.
\item First experimental demonstration for the practical viability of the compact radar system, achieving : 
\begin{itemize}
  \item \textbf{Centimeter-level ranging precision} (RMSE = 1.06 cm) within 1.6–1.9 m range
  \item \textbf{Observable resolution down to 15 cm} for sparse targets under controlled conditions
\end{itemize}
\end{itemize}

The outcomes presented establish fundamental theoretical and experimental foundations for the future integration of atomic-based quantum sensing techniques into practical radar systems. The paper is structured as follows: Section II details the system model of Rydberg atomic homodyne receiver. Section III describes the proposed Rydberg atomic ranging scheme, covering the transmit waveform and signal processing algorithms. Section IV describes the experimental setup and comprehensive validation results and performance analysis. Section V critically discusses system limitations, practical integration challenges, potential applications, and outlines future technological pathways.

\section{System Model}

\begin{figure*}[htbp]
    \centering
    \includegraphics[width=1\textwidth]{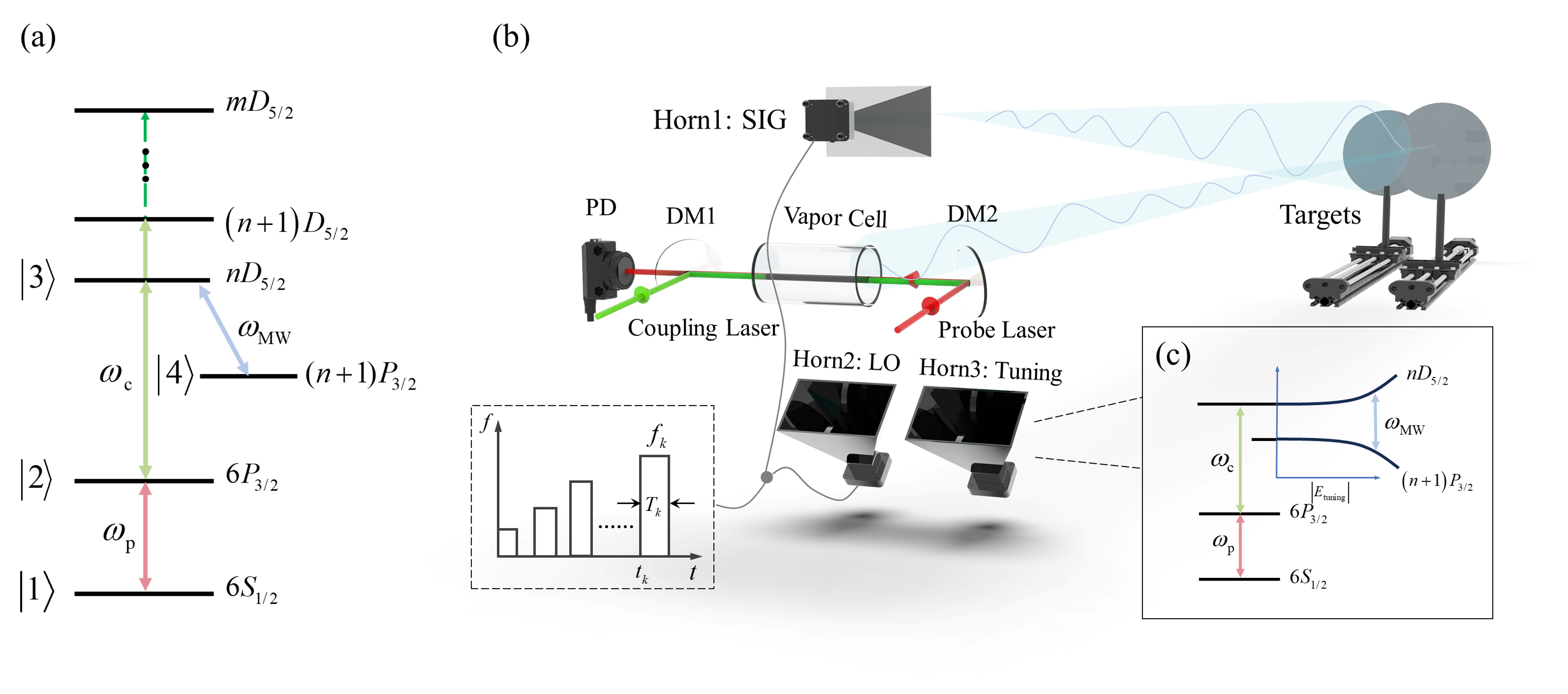} 
    \caption{Energy-level diagrams and ranging methods of the proposed system. A four-level structure enables single-frequency signal reception via EIT and AT splitting. Multi-frequency reception is achieved by switching Rydberg states through laser frequency tuning (a). The frequency hopping of the signal is synchronized with the laser frequency tuning, ensuring that both the echo signal and the LO field are resonantly received and coherently downconverted to a DC optical readout (b). Continuous frequency tunability is realized using an additional far-detuned field (c). DM: dichroic mirror. PD: photodetector.}
    \label{fig:1}
\end{figure*}

The Rydberg atomic homodyne receiver fundamentally relies on a four-level atomic sensing model based on cesium-133 (Cs-133). As depicted in Fig.~\ref{fig:1} (a), the atomic system is driven by two precisely controlled lasers: an 852 nm probe laser excites the transition from the ground state $ |1\rangle = 6S_{\text{1/2}} $ to the intermediate state $ |2\rangle = 6P_{\text{3/2}} $, followed by a 509 nm coupling laser that excites atoms to the Rydberg state $ |3\rangle = nD_{\text{5/2}} $ (or $ nD_{\text{3/2}} $). This configuration establishes an EIT window, which serves as the RF sensing channel. When an external RF field resonantly couples the Rydberg state $ |3\rangle $ to a neighboring state $ |4\rangle $, it induces AT splitting proportional to the RF-field Rabi frequency.

The system dynamics are governed by the Lindblad master equation, formulated as
\begin{equation}
\dot{\rho} = -\frac{i}{\hbar}[H, \rho] + \sum_i \gamma_i \left( L_i \rho L_i^\dagger - \frac{1}{2}\{L_i^\dagger L_i, \rho\} \right),
\label{eq:lindblad}
\end{equation}
where $ L_i $ are collapse operators accounting for spontaneous emission ($ |2\rangle \rightarrow |1\rangle $: $ \gamma_{\text{21}} $) and Rydberg state decoherence ($ |3\rangle $: $ \gamma_{\text{3}} $).
Hamiltonian $H$ is formulated as
\begin{equation}
H = \frac{\hbar}{2} 
\begin{bmatrix}
0 & \Omega_{\text{p}} & 0 & 0 \\
\Omega_{\text{p}} & -2\Delta_{\text{p}} & \Omega_{\text{c}} & 0 \\
0 & \Omega_{\text{c}} & -2(\Delta_{\text{p}} + \Delta_{\text{c}}) & \Omega_{\text{RF}} \\
0 & 0 & \Omega_{\text{RF}} & -2(\Delta_{\text{p}} + \Delta_{\text{c}} + \Delta_{\text{RF}})
\end{bmatrix},
\label{eq:hamiltonian}
\end{equation}
where $\hbar$ is the reduced Planck constant, $ \Omega_{\text{p}} $, $ \Omega_{\text{c}} $, and $ \Omega_{\text{RF}} $ denote the Rabi frequencies of the probe, coupling, and RF fields, respectively. $ \Delta_{\text{p}} $, $ \Delta_{\text{c}} $, and $ \Delta_{\text{RF}} $ represent detunings from respective atomic transitions. For Rydberg atomic homodyne receiver operation, resonant laser excitation ensures $ \Delta_{\text{p}} = \Delta_{\text{c}} = 0 $. 

The RF detection mechanism relies on measuring transmission modulation of the probe laser, proportional to $\mathrm{Im}(\rho_{\text{21}})$, the probe coherence term. For a target at distance $ R $, the echo signal (SIG) field at the atomic vapor cell is
\begin{equation}
E_{\text{RX}}(t) = A E_{\text{SIG}}^0 \cdot e^{j\left(2\pi f_0 t - \frac{4\pi R}{\lambda} + \phi_{\text{SIG}} \right)},
\label{eq:echo}
\end{equation}
where \( E_{\text{SIG}}^0 \) denotes the transmitted signal amplitude, \( A \) represents the round-trip attenuation factor, \( f_0 \) is the carrier frequency, \( \lambda \) the wavelength, and \( \phi_{\text{SIG}} \) the initial phase of the transmitted signal.

A co-located local oscillator (LO) field, generated by a horn antenna at distance \( D \) from the cell, provides the phase reference as
\begin{equation}
E_{\text{LO}}(t) = E_{\text{LO}}^0 \cdot e^{j\left(2\pi f_0 t - \frac{2\pi D}{\lambda} + \phi_{\text{LO}} \right)},
\label{eq:lo}
\end{equation}
where \( E_{\text{LO}}^0 \) is the LO amplitude, \( D \) the fixed LO propagation path length, and \( \phi_{\text{LO}} \) the LO initial phase.

Under the condition \( |E_{\text{SIG}}| \ll |E_{\text{LO}}| \), the probe transmission \( T_{\text{probe}} \) becomes proportional to the magnitude of the beatnote between the echo and LO fields. The total field \( E_{\text{tot}}(t) = E_{\text{LO}}(t) + E_{\text{RX}}(t) \) induces a nonlinear optical response dominated by the LO field. The time-averaged probe transmission is derived as
\begin{equation}
T_{\text{probe}} \propto \lvert E_{\text{tot}} \rvert \approx E_{\text{LO}}^0 + A E_{\text{SIG}}^0  \cdot \cos{\left( -\frac{2\pi (2R - D)}{\lambda} + \Delta\phi \right)},
\label{eq:transmission}
\end{equation}
where the first term corresponds to the LO background, and the second term encodes phase-sensitive target information through the argument of the cosine function. $\Delta\phi$ represents the difference in the initial phases between the SIG field and the LO field.

\section{Proposed Rydberg Atomic Ranging Scheme}
\subsection {Frequency-Hopping Signal Model}
 
The frequency agility of Rydberg atomic homodyne receiver originates from quantum state-selective RF reception governed by Rydberg atom EIT windows. However, the intrinsic narrow instantaneous bandwidth of Rydberg transitions (MHz level) severely limits ranging applications, as the fundamental range resolution depends on the transmitted signal bandwidth

\begin{equation}
\rho_{\text{r}} =  \frac{c}{2B},
\label{eq:range_res}
\end{equation}
where $c$ is the speed of light and $B$ represents the effective bandwidth. This necessitates a frequency-hopping scheme across multiple Rydberg transitions to synthesize GHz-level bandwidth, enabling narrowband reception while achieving broadband-equivalent resolution.

Through selective excitation of Rydberg states $(n, l)$ via laser wavelength tuning (Fig.~\ref{fig:1}(a)), discrete reception bands $f_{\text{34}}^{(0)} \in [f_{\min}, f_{\max}]$ are established, covering DC-THz frequencies. Nevertheless, these response frequencies are inherently fixed and discrete, as they arise from the atomic energy-level structure. To bridge the spectral gaps between these discrete transition points, we employ an AC‐Stark shift compensation method using a far‐detuned control field $E_\text{tuning}$
(see Fig.~\ref{fig:1}(c)). By applying a strong, high‐frequency field that is detuned well away from the Rydberg‐state resonance, the atomic energy levels experience AC‐Stark–induced splitting and shifts. This continuous AC‐Stark tuning of the Rydberg transitions allows the receiver’s response frequency to be swept smoothly across the band, thereby achieving full spectral coverage. From second-order perturbation theory, the effective reception frequency becomes

\begin{equation}
f_{\text{RX}} = f_{\text{34}}^{(0)} + \Delta f_{\text{34}} = f_{\text{34}}^{(0)} + \frac{\alpha_{\text{AC}}}{4h} \lvert E_{\text{tuning}} \rvert^2,
\label{eq:tuned_freq_corrected}
\end{equation}
where $\alpha_{\text{AC}}$ denotes the frequency-dependent AC polarizability, $E_{\text{tuning}}$ is the peak electric field amplitude, and h is Planck's constant. By synergizing discrete Rydberg state jumps (coarse tuning) and continuous AC-Stark shifts (fine tuning), a non-uniform frequency grid $\left\{f_k\right\}_{k=1}^K$ of stepped frequency continuous waveform (SFCW) is synthesized across the S-band as shown in Fig.~\ref{fig:1} (b). The resulting non-uniform stepped frequency signal is expressed as

\begin{equation}
s(t) = \sum_{k=1}^K A \cdot \text{rect}\left(\frac{t - t_k - T_k/2}{T_k}\right) e^{j\left(2\pi f_k t + \phi_k\right)},
\end{equation}
where $\text{rect}(x)$ is Rectangular function ($=1$ if $|x| \leq 0.5$, else $0$). $t_k=\sum_{i=1}^{k-1} T_i$ represents the start time of the $k$-th frequency segment. $T_k$ is the dwell time at $f_k$. Additionally, Non-uniform SFCWs offer advantages over uniform stepping by mitigating range ambiguities and reducing sidelobe levels.

In the reception scheme as described in Eq. (\ref{eq:transmission}), quadrature detection is implemented using two parallel LO paths with \( 0^\circ \) and \( 90^\circ \) initial phase offsets relative to the echo signal. Subtracting the LO-only component \( E_{\text{LO}}^0 \) from \( T_{\text{probe}} \) yields the in-phase (\( I \)) and quadrature (\( Q \)) components as

\begin{align}
I_k &= \kappa_k \cdot A E_{\text{SIG}}^0 \cdot \cos\left( \theta_k \right) + n_I, \label{eq:in_phase_new} \\
Q_k &= \kappa_k \cdot A E_{\text{SIG}}^0 \cdot \sin\left( \theta_k \right) + n_Q, \label{eq:quadrature_new}
\end{align}
where $\theta_k = -{4\pi R}/{\lambda_k} + {2\pi D}/{\lambda_k} + \Delta\phi$ is the phase difference incorporating the target range $R$, the LO path length $D$, and the initial phase offset $\Delta\phi = \phi_{\text{SIG}} - \phi_{\text{LO}}$. Here $\kappa_k$ denotes the frequency-dependent responsivity of the Rydberg receiver (see Eq. (\ref{eq:kappa}), and $n_I$, $n_Q$ represent technical noise in the respective channels.

\subsection {Nonlinear Response}
\label{sec:nonlinear}

The electric dipole moment of Rydberg atoms scales with the square of the principal quantum number ($n^2$), resulting in marked sensitivity variations across distinct spectral regions. This inherent scaling property, when combined with frequency tuning through far-detuned control fields, amplifies sensitivity degradation - particularly manifesting as compromised signal linearity and diminished detection reliability across operational bandwidths. To ensure precision in phase measurement throughout extended frequency ranges, comprehensive calibration of nonlinear responses becomes imperative. The Lindblad master equation Eq. (\ref{eq:lindblad}) provides a rigorous theoretical framework for calculating quantum system response characteristics, theoretically enabling calibration-free operation through direct correlation with fundamental atomic properties. Nevertheless, practical implementation confronts challenges: theoretical computational complexity and practical implementation difficulties. First, the numerical integration of a high‐dimensional density‐matrix model incurs prohibitive computational complexity, and no closed‐form analytical solution exists. Second, the exact atomic, laser, and environmental parameters required for such calculations are difficult to ascertain. Moreover, numerous nonideal effects in practical systems—such as laser power drift, temperature fluctuations, and electromagnetic interference—further degrade the accuracy of nonlinear‐response simulations based on the master equation. To address these challenges, we introduce a novel compensation strategy employing computationally efficient analytical approximations to overcome these limitations, enabling real-time nonlinear compensation while maintaining quantum traceability.

For a simplified cold atomic system under resonance conditions ($\Delta_{\text{p}} = \Delta_{\text{c}} = 0$), the transmitted probe intensity exhibits a characteristic Lorentzian profile \cite{jing2020atomic} formulated as
\begin{equation}
S({\Omega _{{\text{tot}}}}) =  A(1-\frac{{\Gamma^2}}{{\Omega_{{\text{tot}}}^2 + {{\Gamma^2 }}}}),
\label{S}
\end{equation}
where $\Omega _{{\text{tot}}}$ denotes the total Rabi frequency of composite fields, $A$ represents the amplitude of three-level EIT spectroscopy with theoretical expression $A = \alpha \gamma_{\text{21}} \Omega_{\text{p}}/(\gamma_{\text{21}}^2+2 \Omega_{\text{p}}^2)$ , and $\Gamma$ corresponds to the half-width at half-maximum (HWHM) modified for four-level systems written as

\begin{equation}
\Gamma  = \frac{{2\sqrt{2} \Omega _{\text{p}}}}{\sqrt{\Omega _{\text{c}}^2 + \Omega _{\text{p}}^2}}\Gamma_{\text{EIT}}.
\label{eq:gamma}
\end{equation}

The homodyne receiver fundamentally operates by detecting the AT splitting spectrum, where the system gain corresponds to the derivative of Eq. (\ref{S}), formulated as

\begin{equation}
\kappa({\Omega_{\text{tot}}}) = \frac{2A\Gamma^2 \Omega_{\text{tot}}}{{(\Omega_{\text{tot}}^2 + \Gamma^2)^2}},
\label{eq:kappa}
\end{equation}
where $\Gamma_{\text{EIT}} = \left(\Omega_{\text{c}}^2 + \Omega_{\text{p}}^2\right)/\left(2 \sqrt{\gamma_2^2 + 2 \Omega_{\text{p}}^2}\right)$, which represents HWHM of three-level systems. This gain function reaches maximum at $\Omega _{{\text{tot}}} = \Gamma/\sqrt{3}$, defining the optimal LO power. While prior studies \cite{ren2024research,jing2020atomic} extensively discuss LO optimization, practical implementations must account for Doppler broadening in thermal atomic ensembles through spectral corrections. The proposed calibration method integrates analytical approximations with empirical measurements to effectively address this challenge, enabling accurate modeling of the nonlinear response in the vicinity of the optimal LO.

The experimental calibration procedure comprises three key steps: First, we acquire the optimal LO spectrum by identifying the maximal slope point $(\Omega_{\text{max}}, S_{\text{max}})$ on the EIT-AT splitting resonance curve. $\Omega_{\text{max}}$ is determined from the frequency splitting interval $\Delta f$ of the spectral double peaks through the relation $\Omega_{\text{max}} = 2\pi \Delta f$. This approach underlies the SI-traceable electric field calibration of Rydberg atomic sensors based on EIT-AT splitting. Second, we determine the HWHM parameter through $\Gamma = \sqrt{3}\Omega_{\text{max}}$ derived from Eq. (\ref{eq:kappa}) extremum condition. Third, the optimal system gain $\kappa_{\text{max}}$ can be obtained either by pre-scanning the electric field amplitude response of the Rydberg atomic receiver or via heterodyne-based spectral measurements \cite{jing2020atomic}. This value is then used to compute the amplitude parameter as $A=8\kappa_{\text{max}}\Omega_{\text{max}}/3$. By substituting this into Eq. (\ref{S}), a calibrated nonlinear response function is constructed to accurately fit the nonlinear response around the optimal LO, and is expressed as

\begin{equation}
S({\Omega_{{\text{tot}}}}) =  \frac{8\kappa_{\text{max}}\Omega_{\text{max}}{\Omega_{{\text{tot}}}^2}}{3({\Omega_{{\text{tot}}}^2 + 3{\Omega_{\text{max}}^2)}}}+S_{\text{max}}-\frac{2\kappa_{\text{max}}\Omega_{\text{max}}}{3}.
\label{S2}
\end{equation}

To address phase distortions caused by time-delayed signal echoes, the phase difference
 is derived from the I and Q components of both the mixed signal ($S_{\text{I}}$ and $S_{\text{Q}}$) and LO ($S_{{\text{LO\_I}}}$ and $S_{{\text{LO\_Q}}}$). By analyzing the resonant outputs which are expressed as

\begin{align}
I={S^{-1}(S_{\text{I}}) - S^{-1}(S_{{\text{LO\_I}}})},\\
Q={S^{-1}(S_{\text{Q}}) - S^{-1}(S_{{\text{LO\_Q}}})},
\end{align}
where the nonlinear calibration function $S^{-1}(\cdot)$ is defined as the inverse of the nonlinear response $S(\Omega)$. This formulation enables real-time compensation for nonlinear phase shifts across the operational bandwidth, ensuring robust linearization of the sensor response without requiring external references. Experimental validation of this nonlinear compensation framework is presented in Section \ref{results}.

Additionally, Eq. (\ref{eq:gamma}) shows that when the frequency increments $\Delta \Omega_{\text{c}}$ across the operational grid $\left\{f_k\right\}_{k=1}^K$ satisfy the condition $\Delta\Omega_{\text{c}} \ll \Omega_{\text{p}}$, the resulting $\Gamma$ becomes approximately proportional to $\Gamma_{\text{EIT}}$. A single calibration of this proportionality constant suffices to determine optimal LO settings for the entire grid, significantly simplifying multi-frequency operation.

\subsection{Ranging Algorithm for Rydberg Atomic Homodyne Receivers}
\label{sec:ranging_algorithm}

Rydberg atomic homodyne receivers face two technical challenges in practical applications beyond the nonlinear response discussed in Subsection \ref{sec:nonlinear}. First, unlike traditional radar receivers that perform mixing via electronic circuits, the LO field in Rydberg receivers propagates through free space. This spatial propagation may introduce multipath reflections of the LO field itself, causing interference signals to superimpose during the mixing process and degrading coherence. Second, the current Rydberg receiver system noise is dominated by technical noise, including thermal noise from electronic components, laser intensity noise, and mechanical vibration of the measurement system. Some of these noise sources exhibit non-Gaussian characteristics, violating the Gaussian noise assumption in conventional ranging methods.

To address these challenges, we propose a compressive-sensing--centric algorithm that integrates quantum-accurate nonlinear compensation, median filtering, and Huber-loss optimization, hereafter referred to as CS-Rydberg. The Huber regularization is specifically chosen to handle non-Gaussian noise characteristics, while phase normalization focuses the reconstruction on phase information which is critical for ranging accuracy. By exploiting the compressive-sensing paradigm, the proposed CS-Rydberg algorithm achieves accurate reconstruction of sparse range profiles from highly sub-Nyquist frequency samples, thereby substantially reducing the requisite number of sampling points.

\begin{algorithm}[t]
\caption{CS-Rydberg Ranging Algorithm}
\label{alg:cs_ranging}
\begin{algorithmic}[1]
\REQUIRE Non-uniform freq. grid $\{f_k\}_{k=1}^K$ 
\REQUIRE I/Q signals $\mathbf{I}, \mathbf{Q} \in \mathbb{R}^{K \times N_s}$
\REQUIRE LO reference $\mathbf{I}_{\text{LO}}, \mathbf{Q}_{\text{LO}} \in \mathbb{R}^{K \times N_s}$
\REQUIRE Range grid $\mathbf{r} = [r_1, \dots, r_{N_r}]^T$
\REQUIRE Huber param $\delta$, tolerance $\epsilon$
\ENSURE Range profile $\mathbf{p}_{\text{CS}}$

\STATE $\mathbf{s} \gets \text{zeros}(K)$  \hfill \textit{// Initialize signal vector}
\FOR{$k \gets 1$ \TO $K$}
    \STATE $\tilde{I}_k \gets \text{medfilt}(I_k, 50)$  \hfill \textit{// Median filter (impulse noise)}
    \STATE $\tilde{Q}_k \gets \text{medfilt}(Q_k, 50)$
    \STATE $\bar{I}_k \gets \text{mean}(\tilde{I}_k)$  \hfill \textit{// Temporal averaging}
    \STATE $\bar{Q}_k \gets \text{mean}(\tilde{Q}_k)$
    \STATE $\bar{I}_{\text{LO},k} \gets \text{mean}(I_{\text{LO},k})$
    \STATE $\bar{Q}_{\text{LO},k} \gets \text{mean}(Q_{\text{LO},k})$
    
    \STATE \textit{// Nonlinear compensation }($S^{-1}$)
    \STATE $\hat{I}_k \gets S^{-1}(\bar{I}_k) - S^{-1}(\bar{I}_{\text{LO},k})$ 
    \STATE $\hat{Q}_k \gets S^{-1}(\bar{Q}_k) - S^{-1}(\bar{Q}_{\text{LO},k})$
    
    \STATE $s_k \gets \hat{I}_k + j\hat{Q}_k$
    \STATE $\mathbf{s}[k] \gets s_k / |s_k|$  \hfill \textit{// Phase normalization}
\ENDFOR

\STATE $\mathbf{\Phi} \gets \text{BuildMeasurementMatrix}(f_k, r_n, \phi_{\text{ref}}, \phi_{\text{sys}})$  \hfill \textit{// Eq. (\ref{matrix})}
\STATE $\mathbf{x} \gets \text{SolveOptimization}(\mathbf{\Phi}, \mathbf{s}, \delta, \epsilon)$  \hfill \textit{// Eqs. (\ref{eq:opt})}
\STATE $\mathbf{p}_{\text{CS}} \gets |\mathbf{x}|$
\STATE $\mathbf{p}_{\text{CS}} \gets \mathbf{p}_{\text{CS}} / \max(\mathbf{p}_{\text{CS}})$  \hfill \textit{// Normalization}
\RETURN $\mathbf{p}_{\text{CS}}$
\end{algorithmic}
\end{algorithm}

\subsubsection{Signal Preprocessing}
Sequential processing is applied to I/Q signals at each frequency point $f_k$:
\begin{itemize}
    \item \textbf{Median filtering}: Length-50 median filter suppresses impulse noise
    \item \textbf{Temporal averaging}: $N_s$ samples averaged to improve SNR before nonlinear correction
    \item \textbf{Nonlinear compensation}: Apply inverse response $S^{-1}(\cdot)$ (calibrated via Eq. \ref{S2}) to compensate quantum nonlinearity as
    \begin{align*}
        I_k &= S^{-1}(\bar{I}_k) - S^{-1}(I_{\text{LO},k}), \\
        Q_k &= S^{-1}(\bar{Q}_k) - S^{-1}(Q_{\text{LO},k}),
    \end{align*}
    where $I_{\text{LO},k}$, $Q_{\text{LO},k}$ are LO reference components at $f_k$.
    \item \textbf{Phase normalization}: $s_k \gets (I_k + jQ_k)/|I_k + jQ_k|$ eliminates amplitude fluctuations, focusing on phase information crucial for ranging
\end{itemize}

The preprocessing outputs a phase-normalized complex signal vector $\mathbf{s} \in \mathbb{C}^K$.

\subsubsection{Compressive Sensing Reconstruction}
Construct the measurement matrix $\mathbf{\Phi} \in \mathbb{C}^{K \times N_r}$ formulated as

\begin{equation}
\mathbf{\Phi}(k,n) = \exp\left(j\left[\frac{4\pi f_k r_n}{c} + \phi_{\text{ref}} + \phi_{\text{sys}}(f_k)\right]\right),
\label{matrix}
\end{equation}
where $r_n$ ($n = 1,\dots,N_r$) are range grid points, $\phi_{\text{ref}}$ is the fixed reflection phase shift, and $\phi_{\text{sys}}(f_k)$ is the frequency-dependent system transmission delay.

Solve for the range profile via Huber-regularized optimization expressed as
\begin{equation}
\begin{aligned}
\min_{\mathbf{x}} \quad & \sum_{n=1}^{N_r} \phi_\delta(|x_n|),\\
\text{s.t.} \quad & \|\mathbf{\Phi}\mathbf{x}-\mathbf{s}\|_2 \le \epsilon
\end{aligned}
\label{eq:opt}
\end{equation}
where $\phi_\delta(\cdot)$ is the Huber function defined as
\begin{equation}
\phi_\delta(z) = \begin{cases} 
\frac{z^2}{2\delta} & |z| \leq \delta \\
|z| - \frac{\delta}{2} & |z| > \delta ,
\end{cases}
\end{equation}
$\delta$ controls the L1/L2 penalty transition threshold, $\mathbf{x} \in \mathbb{C}^{N_r}$ is the complex range profile, and $\epsilon$ is the noise tolerance.

\subsubsection{Result Postprocessing}
The final output, $\mathbf{p}_{\text{CS}}$, is obtained by normalizing the magnitude of the compressive sensing solution vector $\mathbf{x}$. This normalized range profile enables the resolution of closely spaced targets separated by less than $\Delta r < c/(2B)$, where $B = \max_k f_k - \min_k f_k$ is the total system bandwidth. This sub-wavelength resolution capability fundamentally exceeds traditional limits through the super-resolution inherent in the compressive sensing reconstruction.

\begin{figure*}[htbp]
    \centering
    \includegraphics[width=1\textwidth]{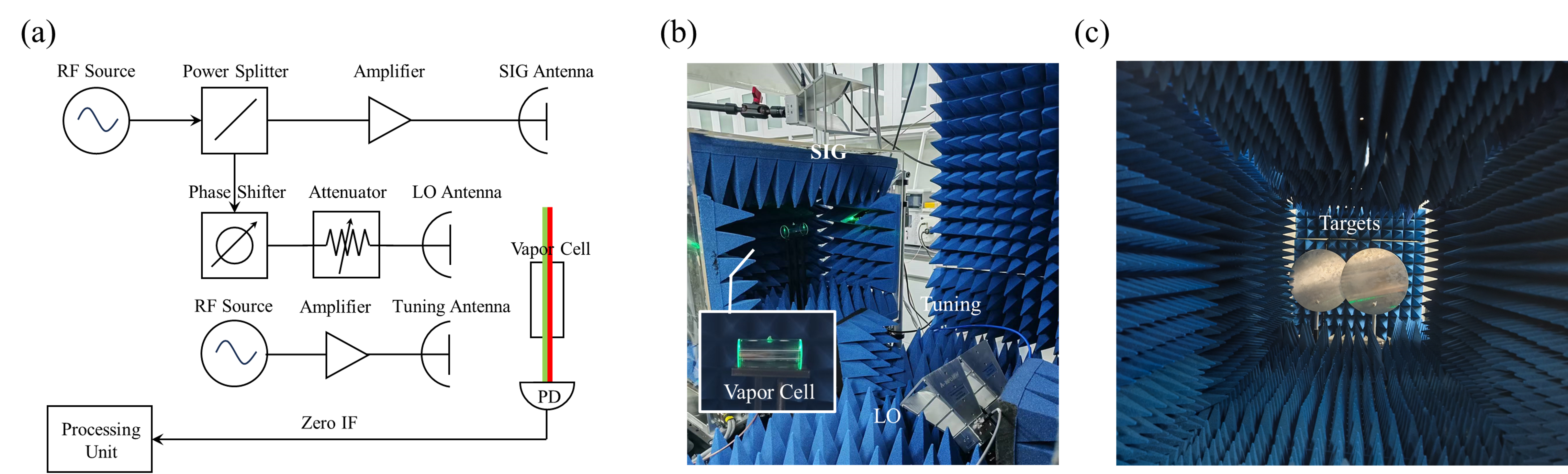} 
    \caption{Overview of radar framework based on Rydberg atomic homodyne receiver (a) and experimental setup (b,c). The SIG field and LO field are generated from the same RF source to maintain long-term phase coherence and power-split. The SIG field is amplified by an amplifier and transmitted toward the target via a horn antenna. The LO field passes through a phase shifter to adjust its initial phase difference relative to the SIG field, and is then optimized in amplitude via an attenuator before propagating toward the atomic vapor cell through another horn antenna. The tuning field is generated by a separate RF source and directed toward the vapor cell. On the receiver side, the probe laser and coupling laser counter-propagate and overlap through the vapor cell, and their transmitted laser is collected by a photodetector and recorded by an oscilloscope. In the experimental setup, the target is a 400-mm-diameter aluminum alloy circular plate mounted on a displacement stage with a 500-mm travel range. }
    \label{fig:radar framework}
\end{figure*}

\section{Experimental Results and Discussion}

\subsection{Experiment implementations}

The core component of the system is a cylindrical Cs-133 vapor cell with a length of \SI{5}{cm} and diameter of \SI{2.5}{cm}. Experiments are conducted in a microwave anechoic chamber to suppress environmental electromagnetic interference and multipath effects. As illustrated in Fig. \ref{fig:1} (b) and Fig. \ref{fig:radar framework} (b,c), a \SI{852}{nm} probe laser excites atoms from the ground state $6S_{1/2}$ to the intermediate state $6P_{3/2}$, with its frequency stabilized via saturated absorption spectroscopy. A \SI{509}{nm} coupling laser further drives atoms to Rydberg states (principal quantum numbers $n = 58$--64). The probe and coupling lasers counter-propagate and overlap within the vapor cell to minimize Doppler broadening. The probe laser operates at a power of \SI{50}{\micro\watt} with a beam waist of $\sim$\SI{1.0}{mm}, while the coupling laser delivers \SI{30}{mW} with a beam waist of $\sim$\SI{1.3}{mm}. Dynamic frequency tuning of the coupling laser is achieved by mechanically adjusting the piezoelectric actuator of the laser's external cavity to selectively match Rydberg transitions corresponding to S-band frequencies. These transitions used in experiments are listed in Table \ref{tab:rydberg_transitions}. The measured transition frequencies are calibrated prior to each experiment and may exhibit slight deviations from nominal values. The maximum frequency interval between adjacent steps is \SI{173}{MHz}, corresponding to a maximum unambiguous radar range of approximately 0.87 m.

\begin{table}[!t]
\caption{Rydberg Transition Frequencies and Measured Parameters}
\label{tab:rydberg_transitions}
\centering
\begin{tabular}{cccc}
\hline
Rydberg & Tuning (+/-)  & Theoretical & Measured \\
States & Field & Frequency (GHz) & Frequency (GHz) \\
\hline

64$D_{5/2}$ & - & 2.625 &  2.640\\
63$D_{5/2}$ & - & 2.757 &  2.766\\
62$D_{5/2}$ & - & 2.899 &  2.912\\
61$D_{5/2}$ & - & 3.050 &  3.063\\
60$D_{5/2}$ & - & 3.212 &  3.225\\
59$D_{5/2}$ & - & 3.386 &  3.398\\
59$D_{5/2}$ & + & / &  3.499\\
58$D_{5/2}$ & - & 3.565 &  3.584\\

\hline
\end{tabular}
\end{table}

The stepped-frequency waveform is synthesized by a vector signal generator (1466H-V, Ceyear) with 8 non-uniform frequency steps. These frequencies are initially calculated using the Alkali Rydberg Calculator (ARC) \cite{vsibalic2017arc} and experimentally calibrated to ensure resonance. Under stable system conditions, the dwell time of each frequency step does not affect one-dimensional range profiling. Thus, dwell times are maximized to ensure frequency and power stability after laser retuning. As shown in Fig. \ref{fig:radar framework} (a), both the SIG field and LO field are generated by the same vector signal generator and split via a power divider to maintain a common time reference and LO coherence over extended periods. The signal path is amplified and transmitted toward the target via a horn antenna. The LO path passes through a voltage-controlled phase shifter to generate in-phase and quadrature components, followed by an attenuator to optimize LO power levels across frequency steps. A far-detuned RF field (\SI{2}{GHz}), generated by a secondary RF source, operates at frequencies far from the Rydberg resonance. High-frequency electric field tuning is adopted to avoid low-frequency shielding effects caused by Rydberg atom adsorption on the vapor cell walls, which would otherwise induce standing waves and phase measurement errors in homodyne detection. This approach also mitigates insertion loss variations from phase shifter and attenuator adjustments. All three RF fields (signal, LO, and detuned fields) employ identical linearly polarized antennas (LB-20180-SF), with the LO polarization aligned parallel to the coupling laser polarization. The horn antenna has a maximum aperture dimension of \SI{13}{cm}, yielding a far-field boundary distance of approximately \SI{23}{cm} at \SI{2}{GHz}. The reason for employing a high-frequency electric field for continuous frequency tuning, rather than a magnetic field \cite{shi2023tunable} or a DC electric field \cite{ouyang2023continuous}, is that devices placed in close proximity to the vapor cell tend to reflect electromagnetic waves. Such reflections exacerbate insertion-loss variations during phase-shifter adjustments, thereby increasing phase-measurement errors. Additionally, the LO field path maintains a distance $\gg$ \SI{45}{cm} from the vapor cell to ensure far-field conditions. The echo signal modulates the Rydberg atomic energy levels, altering the transmission intensity of the probe laser. This perturbation is down-converted to the DC baseband through homodyne detection, captured by a photodetector (PDA36A2, Thorlabs), and recorded using a digital oscilloscope (Tektronix MSO56B) at a sampling rate of \SI{100}{MS/s}. Two aluminum disks (diameter: \SI{400}{mm}) serve as ranging targets, mounted on parallel translation stages to provide adjustable distances. The absolute distance is measured from the antenna aperture plane using a laser rangefinder (\SI{\pm 3}{mm} accuracy). The measured distance consists of three segments: \(R_1\), the distance from the aperture of the SIG horn antenna to the target plate; \(R_2\), the distance from the target plate to the region where the optical beams overlap; and \(D\), the distance from the aperture of the LO horn antenna to the target plate. The target range is then defined as $R = 1/2(R_1 + R_2 - D)$, which cancels the phase shift induced by the vapor‐cell medium and corrects for any offset between the antenna phase center and the mechanical reference.

\begin{figure*}[htbp]
    \centering
    \includegraphics[width=1\textwidth]{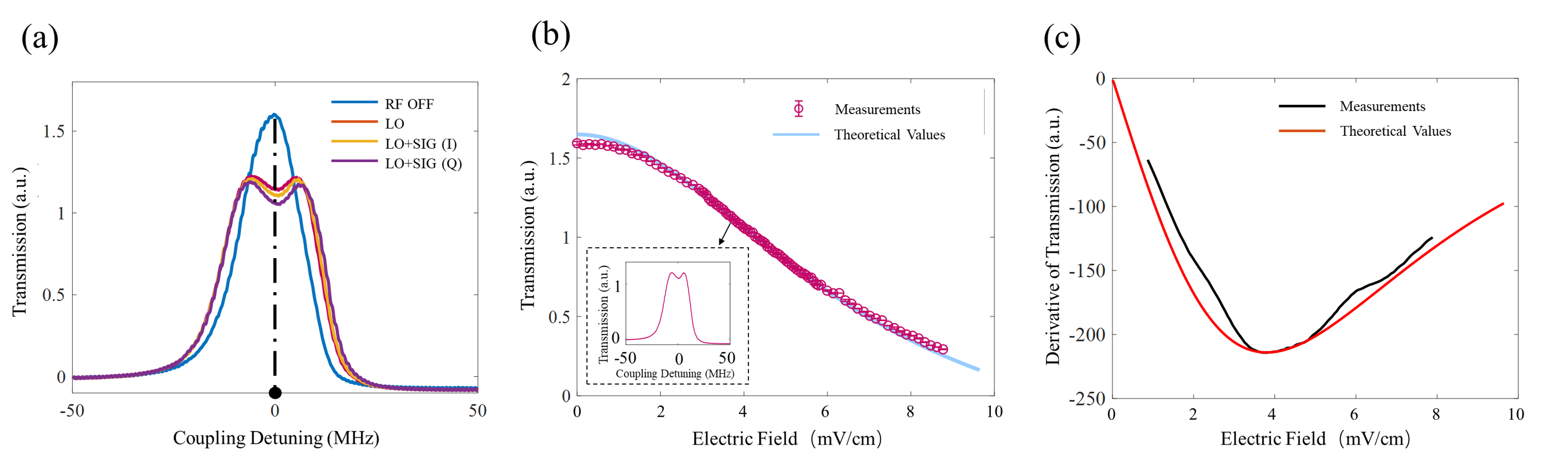} 
    \caption{Experimental demonstration of the Rydberg atomic homodyne receiver and its nonlinear response principle. (a) Spectra under conditions of no RF field (blue curve), LO field only (red curve), in-phase LO and SIG fields (yellow curve), and quadrature-phase LO and SIG fields (purple curve). The transmission value at resonance is taken as the output of the Rydberg atomic receiver. (b) Transmission versus electric field under resonant conditions. Experimental data are shown as red circles. Error bars represent 1$\sigma$ standard errors obtained from 3 measurements. The blue curve shows theoretical calculations based on Eq. (\ref{S2}). The inset displays the AT splitting spectrum under optimal LO conditions and its position in the transmission-electric field relation. (c) Intrinsic gain coefficient of the Rydberg atomic homodyne receiver versus electric field strength, represented by the rate of change of the transmission spectrum. The black curve is the derivative of interpolated experimental data from panel (b) after Gaussian smoothing. The red curve shows the derivative of the theoretical curve in panel (b) according to Eq. (\ref{eq:kappa}).}
    \label{fig:result1}
\end{figure*}

Prior to experiments, the optimal LO power for each frequency step is calibrated by adjusting the attenuator. Resonant frequencies at each LO power level are individually determined experimentally to account for AC-Stark shifts, ensuring optimal alignment of atomic transitions across the frequency steps. Additionally, the orthogonality of the I/Q components is calibrated using a vector network analyzer (VNA) (Ceyear 3657A), recording the phase shifter input voltages corresponding to \ang{0} and \ang{90} phase differences between the LO and SIG fields. To ensure phase stability, electronic components and cables (bending-induced phase variation $\leq$ \ang{0.1}$/${GHz}) are rigidly fixed to minimize mechanical disturbances during operation.

\subsection {Experimental Results}
\label{results}

\begin{figure}[!t]
\centering
\includegraphics[width=\linewidth]{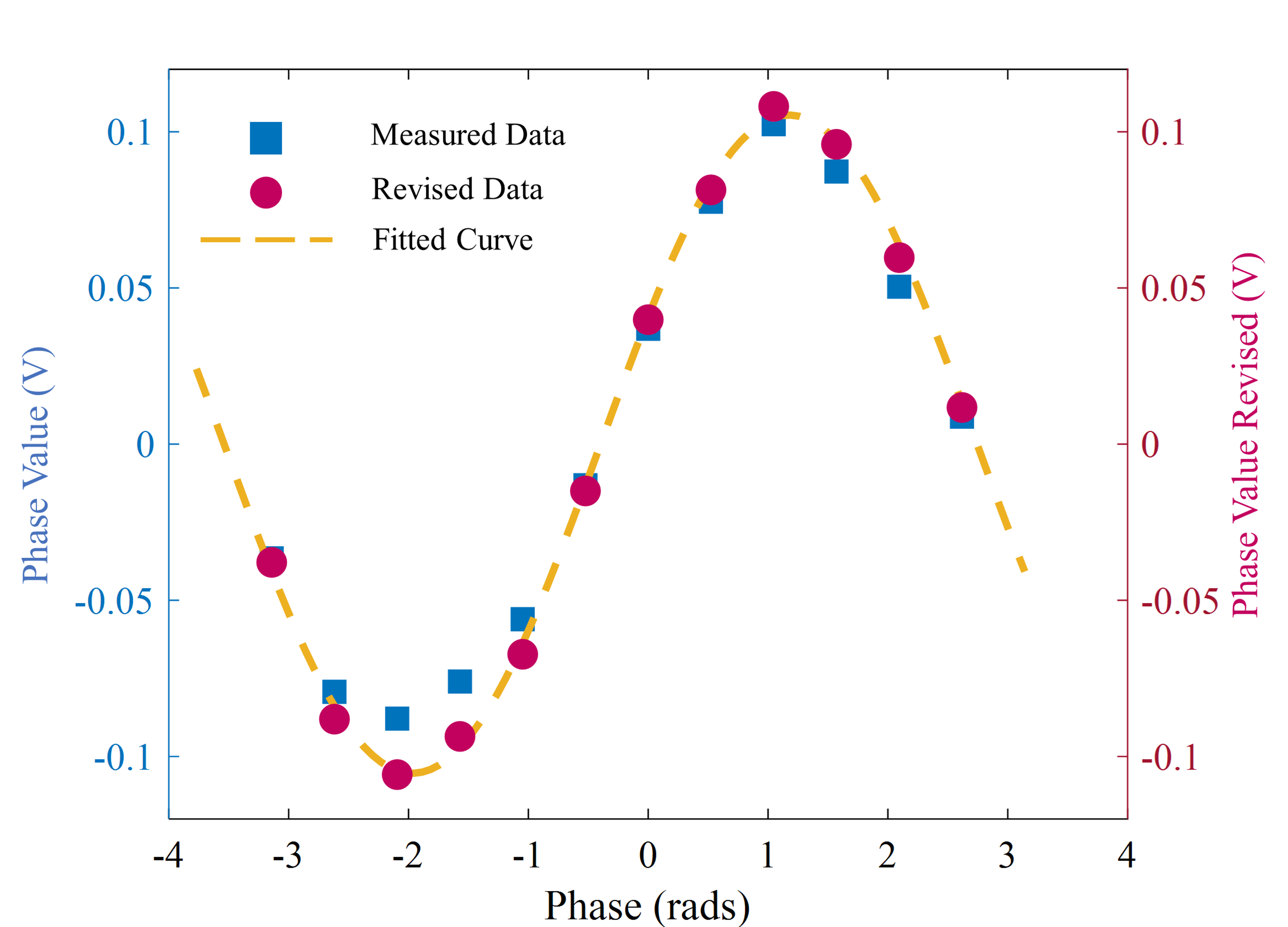}
\caption{Verification of phase measurement capability and orthogonality. Blue solid squares represent the output of the Rydberg atom homodyne receiver with the LO phase adjusted in $\pi/6$ rads increments. Red circles indicate the output revised by the nonlinear response of Eq. (\ref{eq:kappa}). The yellow dashed curve corresponds to a cosine function fitted to the red circles, with $R^2 =99.89\%$.}
\label{fig:phase}
\end{figure}

\begin{figure*}[htbp]
    \centering
    \includegraphics[width=1\textwidth]{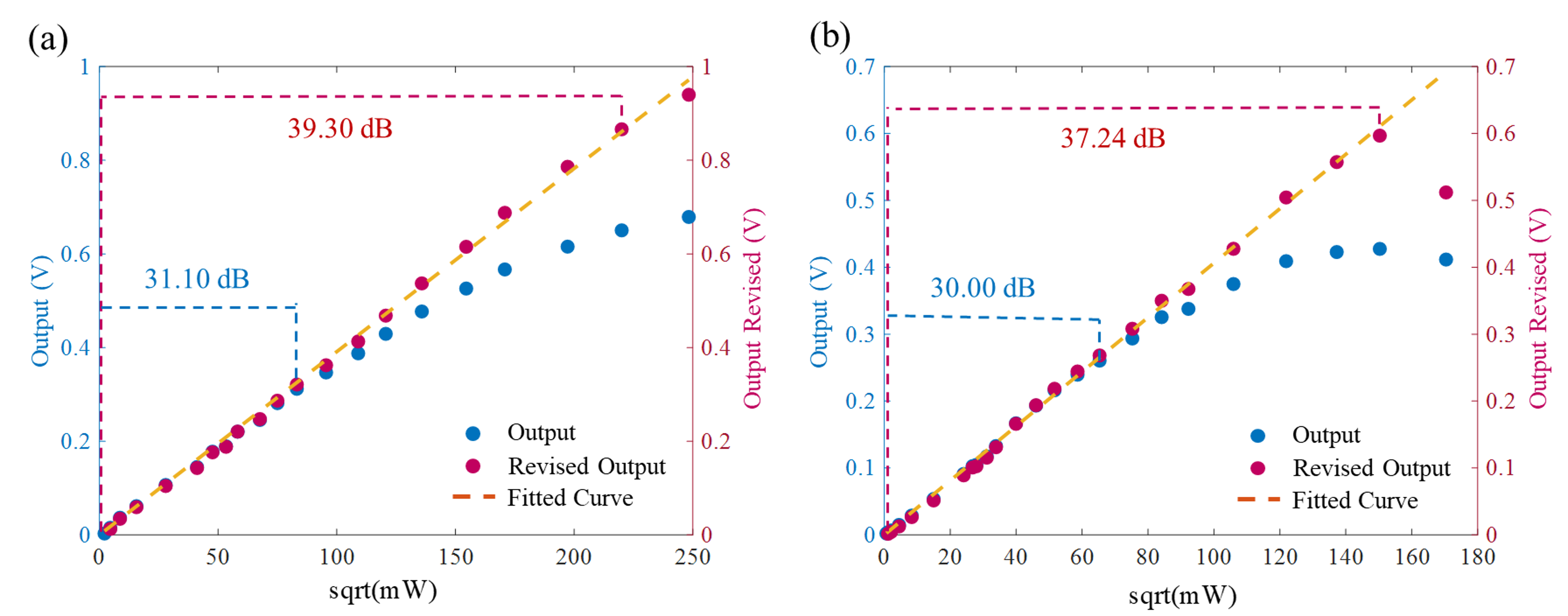} 
    \caption{Linearity and dynamic range of the Rydberg atomic homodyne receiver under (a) coherent enhancement and (b) coherent cancellation conditions. Blue circles represent the raw demodulated output; red circles represent the output after nonlinear compensation; the yellow dashed line is a linear fit to the small-signal regime. In (a), the uncorrected linear dynamic range is 31.10 dB, extending to 39.30 dB after compensation. In (b), it is 30.00 dB uncorrected and 37.24 dB with compensation. In both cases, a dynamic range improvement of over 7 dB is achieved after nonlinear compensation.}
    \label{fig:linearity}
\end{figure*}

\begin{figure}[!t]
\centering
\includegraphics[width=\linewidth]{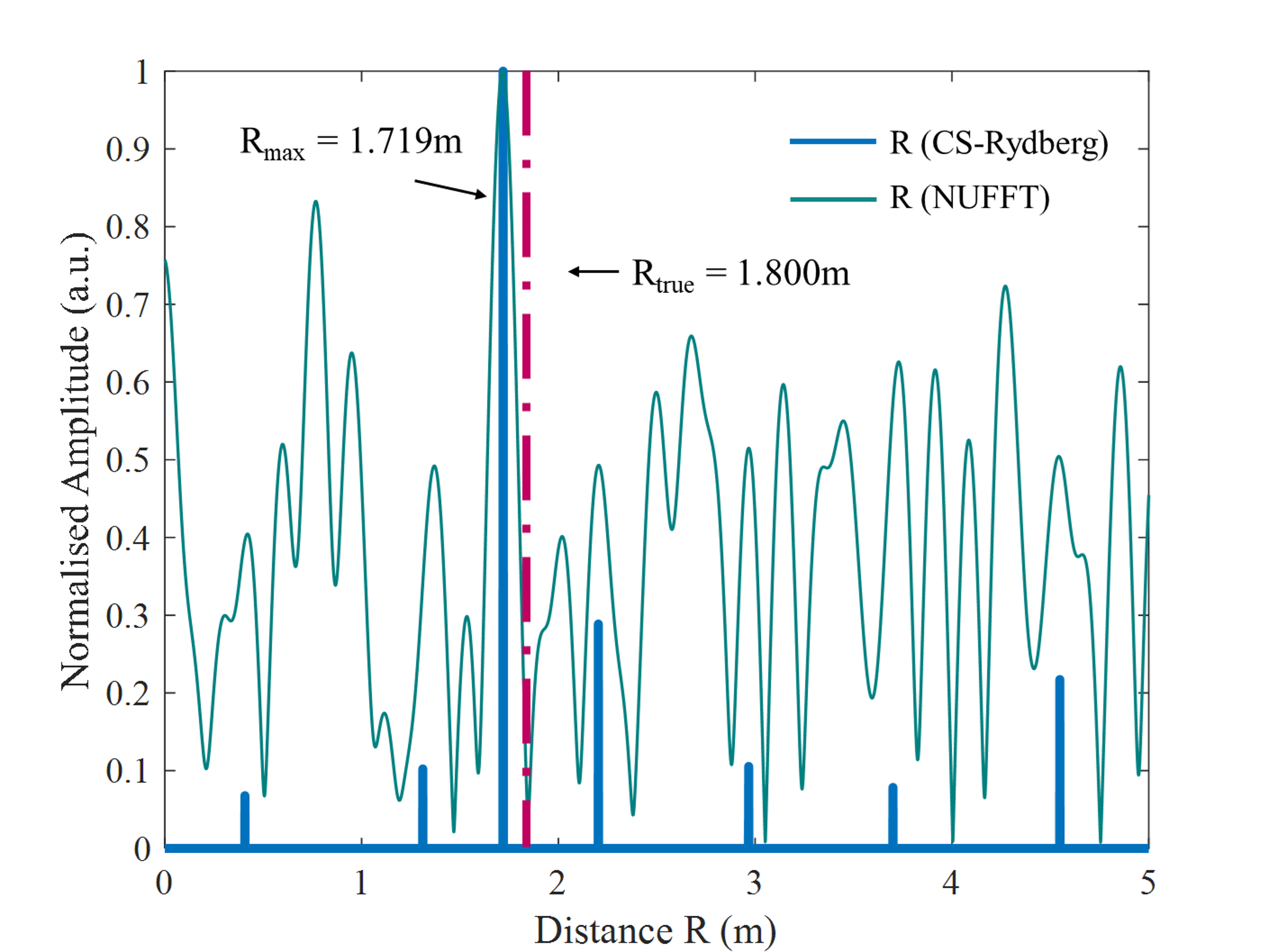}
\caption{Single-target absolute distance measurement. The green curve represents the one-dimensional range profile reconstructed via non-uniform FFT (NUFFT), while the blue curve represents the profile reconstructed via CS-Rydberg. The horizontal axis denotes range bins, and the vertical axis denotes the normalized echo amplitude at each bin. The red dashed line indicates the target distance measured by a laser rangefinder.}
\label{fig:single}
\end{figure}

\begin{figure*}[htbp]
    \centering
    \includegraphics[width=1\textwidth]{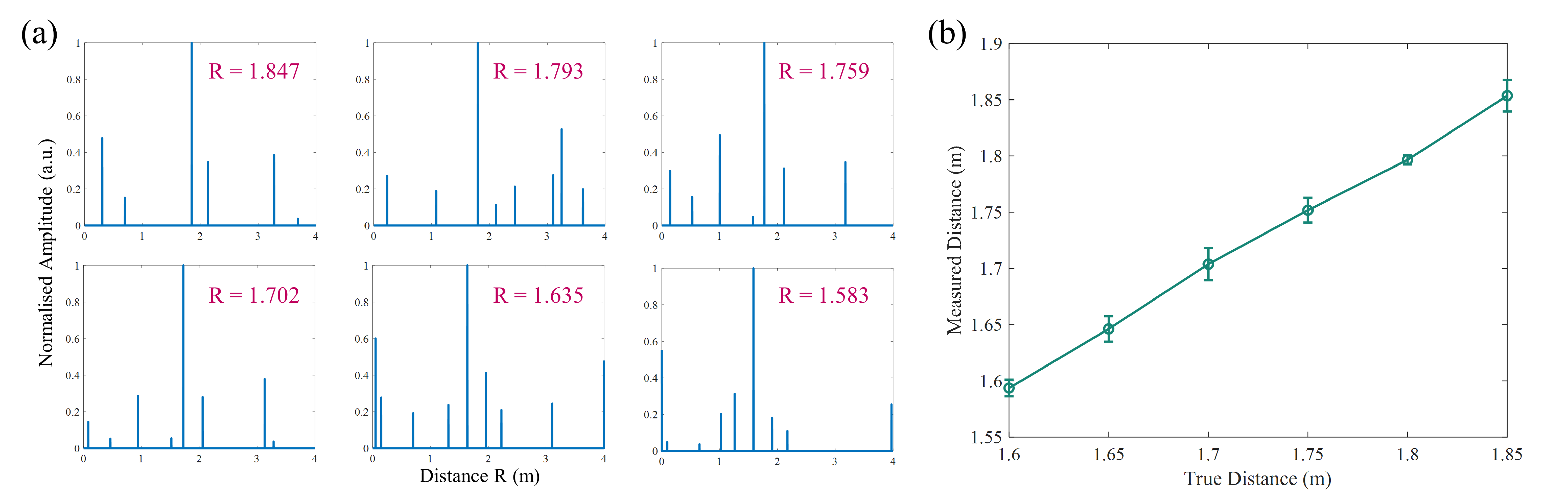} 
    \caption{Single‐target relative displacement and ranging repeatability experimental results. The target plate is translated on a motorized stage from 1.60 m to 1.90 m in 0.05 m increments. The measurement at 1.90 m is used for single‐point calibration and to align the remaining distances. (a) Ranging results for six target positions in a single measurement run. (b) Summary of five independent measurement runs. The horizontal axis denotes the true target distance, and the vertical axis shows the ranges measured by the radar system based on Rydberg atomic homodyne receiver. Error bars represent ±1 $\sigma$ standard error from five measurements. The system achieves a root mean squared error (RMSE) of 1.06 cm and maximum measurement deviation of 2.3 cm.}
    \label{fig:rangings}
\end{figure*}

Fig. \ref{fig:result1} (a) shows the measured probe transmission as a function of the applied RF electric field under four conditions: (i) no RF, (ii) LO only, (iii) signal + LO in-phase, and (iv) signal + LO in quadrature. The transmission curves highlight the response of the Rydberg receiver. In the absence of any RF (no RF), the probe laser transmission remains at its baseline level (the natural EIT transmission with no perturbing field). With only the LO field applied, the transmission shifts due to the AT-splitting from the LO, establishing a bias point on the atomic resonance. When an echo signal is added in-phase (0° out of initial phase with the LO), the total RF field amplitude increases. Conversely, adding the signal in quadrature (90° out of initial phase with the LO) produces a larger net change in transmission. This is because, under these conditions, the echo signal undergoes stronger cross‐interference with the local oscillator, resulting in a larger perturbation of the atomic medium and a more pronounced change in transmission. The in-phase and quadrature outputs can be used to reconstruct the echo signal incident on the atomic vapor cell. Fig. \ref{fig:result1} (b) shows the variation of the spectral transmission under resonant conditions as a function of the applied electric field strength. The experimental data (markers) are in excellent agreement with the theoretical EIT-based model (solid curves), which predicts a saturating response as the RF field increases (per Equation \ref{S2} in the Theory section). Deviations between experiment and theory grow progressively at very low and very high fields, owing to AC–Stark–induced shifts in the resonance frequency, which detune the probed transition from resonance. The inset depicts the spectrum recorded at the optimal LO bias. Compared with the results reported in \cite{jing2020atomic, yang2023amplitude}, the corresponding field values here are smaller, likely due to additional Doppler-broadening effects. Crucially, the derivative of the transmission with respect to the electric field, also shown in Fig. \ref{fig:result1}(c), illustrates the receiver’s gain. This derivative (d$T$/d$E$) peaks at an intermediate field amplitude – indicating an optimal operating point where the atomic medium is most sensitive to small changes in $E$. At low fields, the slope is initially small (the atoms are in the linear regime of response). As the LO bias raises the operating point, the slope increases, reaching a maximum gain where the system responds most strongly to the signal. Beyond this point, the slope decreases as the Rydberg atoms saturate: additional field produces diminishing changes in transmission. This gain saturation is a direct consequence of the atomic population approaching the upper state limit (once a significant fraction of atoms are already excited by the combined LO+SIG field, the transparency can no longer change linearly with further field increase). Physically, this behavior mirrors a classical receiver’s amplifier saturation – here manifested in the optical domain via EIT. From this fit, one can extract an estimated maximum incremental gain of the atomic receiver – for example, the peak derivative corresponds to a gain of approximately 213 a.u. of transmission per V/m (arbitrary units), achieved at an RF field around $E\approx 3.7$ mV/cm (just as an illustrative number; actual scale depends on atomic parameters). This confirms that the receiver operates optimally in a certain field range; outside that range, sensitivity drops.

Having established the receiver’s amplitude response, we next verify its ability to measure RF phase accurately. In a homodyne system, the output should depend on the phase difference between the signal and LO as $V_{\text{out}} \propto \cos(\Delta\phi)$ (per Equation \ref{eq:transmission} for the in-phase channel, with the quadrature channel yielding a $\sin(\Delta\phi)$ dependence). To test this, the LO phase is incrementally shifted in steps of $\pi/6$  while keeping the signal amplitude constant. For each relative phase, the receiver’s output (optical transmission change) is recorded. Because the amplitude response of the atomic sensor is nonlinear (as seen in Fig. \ref{fig:phase}), a calibration curve or nonlinear compensation is applied to convert the raw transmission changes into a linearized output proportional to the effective RF field. Fig. \ref{fig:phase} plots the normalized receiver output as a function of the LO phase shift after this compensation. The data points lie almost perfectly on a cosine function (shown as a solid fitted curve). The cosine fit has an  coefficient of determination $R^2 = 99.89\%$, indicating an almost perfect correspondence with the expected $\cos$ relationship. This demonstrates two important aspects: (1) the Rydberg receiver preserves the phase information of the RF signal with high fidelity, and (2) the in-phase and quadrature components of the signal are properly orthogonal and balanced. In other words, when the LO phase is swept, the in-phase channel output varies cosinusoidally without leakage of any quadrature component, and a 90° phase shift in the LO indeed moves the receiver output through the full sinusoidal cycle (indicating the presence of a corresponding quadrature channel if measured). The near-unity $R^2$ and the absence of distortion in Fig. \ref{fig:phase} confirm that the atomic mixer behaves as an ideal IQ demodulator for small signals: it coherently converts the RF field’s phase into a measurable optical output. Any slight deviations from the perfect cosine (the data points are almost on top of the fit curve, with maximum deviations $< 1\%$) are attributable to experimental uncertainties such as residual amplitude modulation or imperfect compensation of the nonlinear response. Importantly, these results verify that phase coherence is maintained in the atomic detection process – the LO and signal interference inside the vapor cell yields a stable and repeatable phase-dependent output. This capability is crucial for applications like interferometric sensing or coherent radar, as it means the Rydberg receiver can function analogously to a conventional coherent detector, providing both amplitude and phase of the RF field.

Fig.~\ref{fig:linearity} evaluates the receiver’s linearity and dynamic range by plotting the demodulated output \(V_{\mathrm{out}}\) versus equivalent input voltage under two phase conditions: (a) coherent enhancement of the SIG field with the LO field, and (b) coherent cancellation. In both cases, the LO field is held at a high-sensitivity bias point, while the SIG field—calibrated by a vector network analyzer—is swept from the system noise floor up to saturation. In the small-signal regime, both the raw output and the predistorted output lie on an yellow dashed linear fit, confirming $ V_{\mathrm{out}}\approx \kappa\,E_{\mathrm{sig}}$. Beyond this region, the slope deviates from the reference line due to atomic-saturation-induced compression, and at the highest inputs the output nearly plateaus. Quantitatively, in the coherent-enhancement case the uncorrected linear dynamic range spans approximately 31.10 dB, while after applying nonlinear compensation it extends to 39.30 dB, a gain of 8.20 dB. In the coherent-cancellation case, the raw range is about 30.00 dB, increasing to 37.24 dB with compensation—an 7.24 dB improvement. Thus, compensation enlarges the usable dynamic range by over 7 dB in both scenarios, covering more than two orders of magnitude in input amplitude. The lower limit of each range is set by the system noise floor—dominated by laser noise, detector noise, and atomic projection noise—which defines the minimum detectable signal.

\begin{figure}[!t]
\centering
\includegraphics[width=\linewidth]{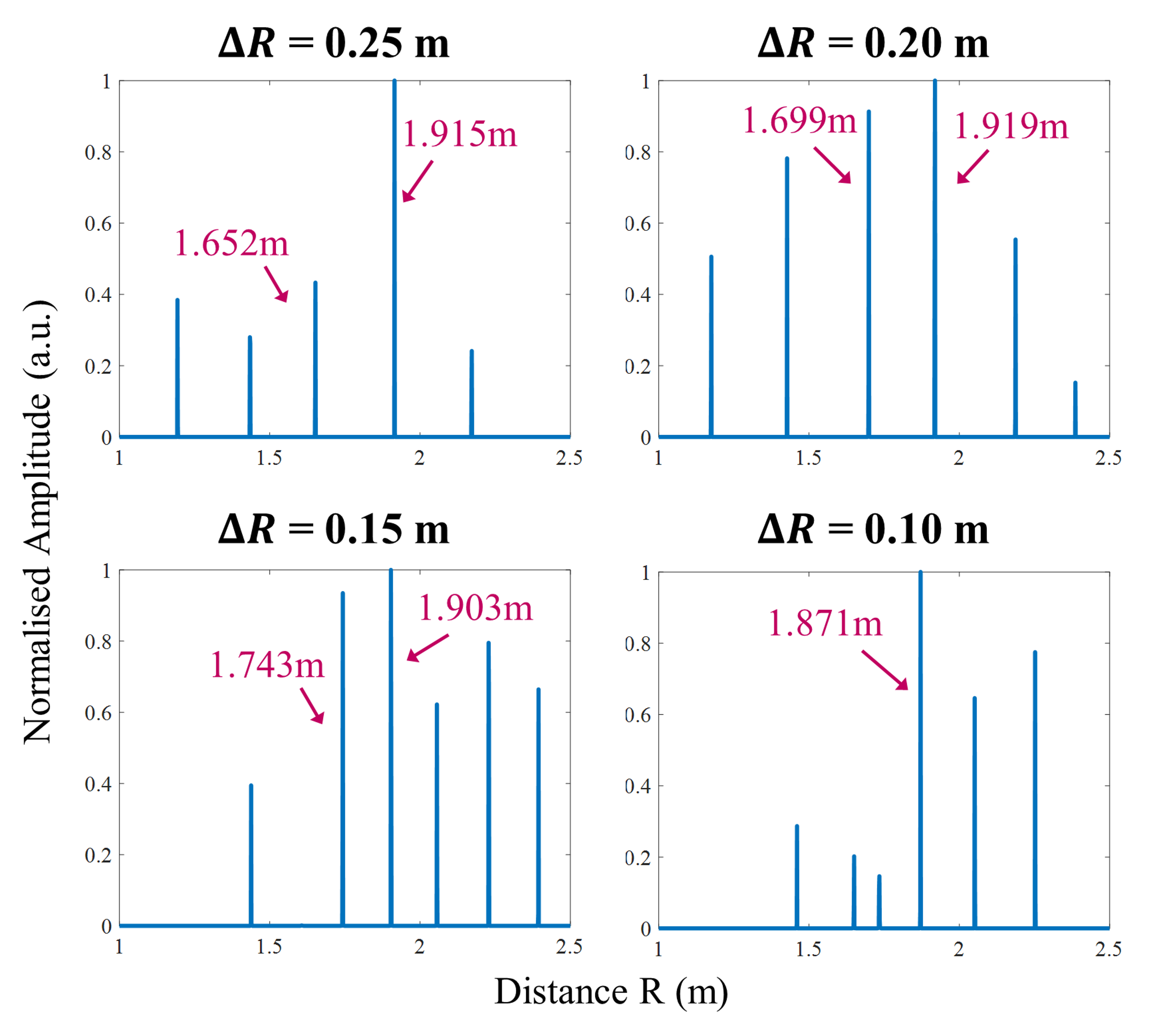}
\caption{Dual‐target resolution experiment results. Two target plates are placed within the transmit beam and mounted on motorized stages: one plate fixed at 1.90 m, the other translated from 1.65 m to 1.80 m in 5 cm increments. When restricting the ranging window to 1-2.5 m (to mitigate phase wrapping effects), two distinct peaks corresponding to the true target positions are observed for separations of 25, 20 cm, and 15 cm. For separations = 10 cm, the peaks merge and are not consistently resolvable under sparsity constraints. This indicates an achievable target separation of 15 cm in sparse scenarios.}
\label{fig:doubletargets}
\end{figure}

After characterizing the receiver’s baseband performance, radar ranging experiments are conducted to validate its practical capabilities. Utilizing the Rydberg atomic homodyne receiver as an RF front-end, we establish a simple bistatic radar configuration. An RF source emits a probing signal, which reflects from the target and is subsequently captured by the Rydberg atomic receiver. In the following, all reported measured and actual distances represent the averaged two‐way path lengths of the target’s transmitted and echo signals after processing, with the propagation distance of the LO field subtracted. Fig. \ref{fig:single} presents the obtained one-dimensional range profile for a single-target scenario. In this experiment, a metal plate is fixed at a known distance of approximately 1.80 m from the Rydberg sensor. The range profile plots the processed receiver output as a function of distance. Fig. \ref{fig:single} (green line) shows the one-dimensional range profile generated by a Non-Uniform Fast Fourier Transform (NUFFT) method. The width of the echo peak represents the system’s range resolution, defined as the minimal distinguishable distance between two targets, inversely proportional to the signal bandwidth (as expressed in Eq. \ref{eq:range_res}). The theoretical resolution corresponding to the applied RF chirp bandwidth is about 15 cm, consistent with our experimental observations. However, the main echo peak appears at around 1.719 m, significantly deviating from the target's known physical position. The discrepancy primarily arises from the effective phase-center drift caused by the finite size of the reflective metal plate, cumulative path measurement errors, and polarization mismatch induced by the angular deviation of the horn antenna. Additionally, several spurious peaks appear in the profile, mainly due to limited frequency sampling points, target distances exceeding the system's maximum unambiguous range, and residual multipath effects. Consequently, we implement the CS-Rydberg algorithm to accommodate undersampling conditions and enhance robustness, as illustrated by the blue line in Fig. \ref{fig:single}. The CS-Rydberg algorithm effectively mitigates undersampling artifacts and multipath interference, significantly enhancing the clarity and robustness of target detection compared to conventional NUFFT.

Therefore, under the current experimental conditions, it is not possible to validate the capability of Rydberg atoms to perform direct ranging based on absolute phase measurements. Instead, ranging is achieved by performing a one-time calibration to determine the relative positional offset. We evaluate the accuracy and repeatability of measuring relative displacement of a target. A metallic reflector is moved across seven known distances ranging from 1.60 m to 1.90 m in 5 cm increments. At each position, five independent ranging measurements are performed using the Rydberg atomic homodyne receiver. To eliminate systematic offset in absolute distance estimation, all measured ranges are uniformly shifted such that the echo corresponding to the 1.90 m reference target (averaged over 3 measurements) aligns precisely with its true position. Fig. \ref{fig:rangings} (a) illustrates the one-dimensional range profiles from one independent set of measurements across the six target positions (excluding the calibration point at 1.90 m). Each measurement consistently exhibits a clear echo peak at the known target distances. Fig. \ref{fig:rangings} (b) displays the average measured distance against the actual distance, with each data point representing the mean of five trials and error bars indicating ±1 standard deviation. Results show exceptional measurement accuracy, consistently aligning with true distances within minimal deviations—less than 3 cm across the entire tested range (1.60–1.85 m). The mean error between the measured and true distances is –0.08 cm, indicating that the system exhibits negligible systematic bias. The root-mean-square error (RMSE) of the measurements is 1.06 cm. Such precision aligns with expectations based on the system’s temporal resolution capabilities, confirming no significant hidden errors in the ranging process. Practically, this confirms that the Rydberg atomic receiver can reliably measure target distances at centimeter-level accuracy within short ranges. The high repeatability also underscores the excellent consistency of system performance throughout the measurement duration. Thus, the data in Fig. \ref{fig:rangings} strongly supports the feasibility of employing quantum sensors in realistic remote sensing applications, confirming no compromise in measurement capability due to the atomic medium.

To experimentally verify the Rydberg atomic radar’s range resolution capability, we conduct a dual-target resolution test using two metallic reflectors. One target is fixed at approximately 1.90 m, and the other is mounted on a movable stage allowing precise distance adjustments relative to the radar sensor. Initially, the movable reflector is positioned at 1.65 m, establishing a 25 cm target separation. Subsequently, the second target is incrementally moved closer, from 1.65 m to 1.80 m in 5 cm increments, recording the resulting one-dimensional range profiles reconstructed using the CS-Rydberg algorithm at each step. Due to the sparse frequency points under current experimental conditions and the more severe multipath effects with two targets, many false peaks appear in the range profile. Nevertheless, gradually approaching peaks corresponding to both targets can still be observed, as shown in Fig. \ref{fig:doubletargets}. When target separations are 15 cm or greater, two distinct and clearly separable peaks consistently appear, corresponding to each reflector. However, as the separation reduces below 15 cm, the two echoes merge into a single combined response, rendering the algorithm unable to distinctly resolve both targets. This suggests that 15 cm represents the minimum observable separation in this experiment, though statistical validation is required to establish a robust resolution threshold. The observed limit (~15 cm) approaches the theoretical resolution of 16 cm enabled by the synthesized 1-GHz bandwidth. Sparse sampling restricts the practical resolution from reaching the theoretical optimum, while the CS-Rydberg algorithm mitigates sidelobe interference to preserve target visibility. Critically, this demonstrates the Rydberg receiver’s ability to resolve closely spaced targets within operational constraints.

\subsection {Discussions and Future Prospects}

The Rydberg atomic receiver permits atomic energy-level tuning via laser frequency adjustment, enabling quasi-continuous frequency coverage from DC to THz—far exceeding conventional radar tunability. However, frequency agility faces inherent limitations: switching operational frequencies requires laser relocking and thermal restabilization. This severely constrains real-time target tracking and dynamic scene response. Current latency is orders of magnitude higher than electronic systems' microsecond-scale beam switching capabilities. Emerging rapid-retuning laser technologies may address atomic radar systems' future requirements for high-dynamic target tracking or high-throughput imaging \cite{stern2020broadly}.

Limited by the scale of the darkroom and laser performance, we are unable to achieve greater equivalent bandwidth and more frequency-domain sampling. In this work, the quality of range imaging is inevitably affected by multipath interference and undersampling. Consequently, only the functional feasibility of the system can be verified, while it fails to test the system’s ultimate capabilities. Nevertheless, the significant potential of Rydberg atomic radar in multiple application scenarios can be inferred. Its inherent quantum traceability and absolute calibration capabilities make it particularly suitable as a reference receiver in high-precision electromagnetic field metrology—for example, serving as a calibration standard for RF antenna testing or radar cross-section (RCS) measurements. Additionally, the Rydberg receiver’s natural immunity to electronic saturation effects and its extremely low thermal noise floor render it highly appealing for passive sensing in electromagnetically silent environments, such as space-based platforms operating in orbit, astronomical observation sites, or covert monitoring scenarios.

A number of key technologies need to be further investigated to realize the potential of these applications. It is possible to evolve from zero IF to heterodyne detection schemes. This only requires introducing a small frequency offset in the reference field to convert the signal to IF thus avoiding the baseband 1/f noise and DC offset artifacts. This also avoids the oscillator pulling effects, which accumulates errors in the actual phase measurement due to the actual gain variation. Consideration is also given to controlling both fields with phase-stable references, which requires advanced signal generation equipment.

In addition, more technological breakthroughs in compactness are required for the system to evolve from the laboratory to practical deployment. Although research has been done to develop chip-scale vapor chambers with integrated photonic devices, maintaining atomic coherence times, optical stability in the external environment is still the challenge. Careful design of thermal management, magnetic shielding, and optical alignment schemes are needed to achieve detection performance comparable to that of laboratory desktop configurations within millimeter-scale chambers.

The ranging system in this study is a foundational validation and a key consideration for the realization of future multi-cell systems such as the Rydberg Atomic Phased Array. Such configurations can provide angular resolution and beamforming capabilities comparable to electronic MIMO radars, and their realization requires precise channel-to-channel amplitude/phase matching, gain drift compensation, and optical path length stability. The Rydberg Atomic Receiver has a natural advantage of expandability due to its size and optical readout characteristics. A shared laser source can be distributed to multiple vapor cells via fiber optics. Large vapor cells with spatially multiplexed optical readout paths can also be designed. Deep consideration of phase coherence maintenance and timing synchronization challenges is required in engineering practice. Deep learning algorithms have the potential to enhance the calibration and tuning of atomic vapor chamber arrays.

\section{Conclusion}

In this paper, we introduce and experimentally demonstrate a transformative radar ranging methodology based on Rydberg atomic homodyne detection, addressing fundamental challenges faced by conventional radar front-ends. The atomic receiver architecture capitalizes on quantum-optical interactions within cesium vapor cells to achieve RF-to-optical signal conversion without traditional RF components such as mixers, amplifiers, and filters. By employing a stepped-frequency approach combined with laser frequency tuning and AC-Stark shift techniques, we effectively synthesize an equivalent GHz-level bandwidth from intrinsically narrow atomic resonances, enabling centimeter-level precision ranging. In this work, we develop a nonlinear response model for a Rydberg atomic homodyne receiver and, by applying a nonlinear compensation scheme, which extends its linear dynamic range by over 7 dB. Furthermore, we introduce a dedicated CS-Rydberg signal-processing paradigm based on compressive sensing. Our experimental results demonstrate that the atomic radar achieves a ranging accuracy of RMSE = 1.06 cm and resolvable target separations down to 15 cm are observed under controlled sparse scenarios.

Despite these advancements, the current homodyne architecture encounters multipath interference and LO leakage, which are experimentally observed as spurious echoes and phase instabilities in measured range profiles. The practical frequency agility is also constrained by laser tuning speed and atomic coherence limitations, currently restricting real-time responses in rapidly changing scenarios. Future research directions include transitioning to a superheterodyne architecture to circumvent DC noise and enhance Doppler detection sensitivity and dynamic range, developing rapid-frequency-switching capabilities to overcome current response delays, and miniaturizing atomic cells and optical components for scalable integration into operational platforms such as UAVs and satellites.

Overall, the presented results strongly support the practical viability and future potential of quantum sensing based on Rydberg atomic receivers. This study provides critical experimental validation and theoretical groundwork, paving the way for revolutionary advancements in precision remote sensing, electromagnetic calibration, and secure sensing operations in next-generation radar systems.

\bibliographystyle{IEEEtran} 
\bibliography{Bibliography} 

\vfill

\end{document}